\documentclass[twocolumn,prb,aps,amsmath,amssymb,floatfix,showpacs]{revtex4-1}

\usepackage{color}
\usepackage{graphicx}% Include figure files
\usepackage{dcolumn}% Align table columns on decimal point
\usepackage{bm}% bold math
\usepackage{subfigure}
\begin{document}

\title{Density-functional calculation of static screening
in 2D materials: \\the long-wavelength dielectric function of graphene}

\author{Thibault Sohier}
\author{Matteo Calandra}
\author{Francesco Mauri}

\affiliation{Institut de Min\'eralogie, de Physique des Mat\'eriaux, 
et de Cosmochimie (IMPMC), Sorbonne Universit\'es - UPMC Univ Paris 06, 
UMR CNRS 7590, Mus\'eum National d'Histoire Naturelle, IRD UMR 206, 
4 Place Jussieu, F-75005 Paris, France.}

\date{\today}% It is always \today, today,

\begin{abstract}

We calculate the long-wavelength static screening properties 
of both neutral and doped graphene in
the framework of density-functional theory.
We use a plane-wave approach with periodic images in the third dimension
and truncate the Coulomb interactions to eliminate spurious interlayer 
screening.
We carefully address the issue of 
extracting two dimensional dielectric properties from 
simulated three-dimensional potentials. 
We compare this method with analytical expressions derived for 
two dimensional massless Dirac fermions in the random phase
approximation. 
We evaluate the contributions of 
the deviation from conical bands, exchange-correlation and local-fields.
For momenta smaller than twice the Fermi wavevector,
the static screening of graphene within the 
density-functional perturbative approach agrees with the
results for conical bands within random phase approximation
and neglecting local fields.
For larger momenta, we find that the analytical 
model underestimates the static dielectric function by $\approx 10\%$, 
mainly due to the conical band approximation.

\end{abstract}

\maketitle

%\tableofcontents

\section{Introduction}
\label{sec:Intro}

The electronic properties of two-dimensional (2D) materials have been
intensively studied in the past decade. They offer the opportunity to probe 
exciting new low-dimensional physics, as well as promising prospects in 
electronic device applications.
Among those interesting properties is electronic transport. 
In the context of electronic transport in graphene, screening is crucial
for electron scattering by charged impurities 
\cite{Katsnelson2006,Fogler2007,Hwang,Hwang2009}, 
electron-phonon coupling \cite{VonOppen2009,Sohier2014a,Fratini2008}, or 
electron-electron interactions \cite{Kotov2008}.

Dimensionality is well known to be essential in determining 
the physical properties of materials. Correctly describing 
the physics of 2D materials requires careful modeling and definition of the 
relevant physical quantities. This is particularly true for {\it ab initio}
calculations based on plane-wave basis set, as they rely 
on periodic boundary conditions along the three dimensions.
In this framework,  when simulating low-dimensional materials, 
periodic images of the system are necessarily included in the calculation. 
For some physical properties, 
the interactions between the periodic images are sufficiently 
suppressed by imposing large distances between them\cite{Lebegue2009}. 
However, if the electronic density is perturbed at small
wavevector, long-range Coulomb interactions between electrons 
from different periodic images persist even for very large distances,
leading to some spurious screening.
On the other hand, {\it ab initio} calculations have
the advantage of describing a complete band structure and 
accounting for local fields.
Local fields designate electronic density perturbations
at wavelengths smaller than the unit-cell dimensions 
\cite{Adler1962,Pick1970,Sinha1974}.
Accounting for local fields usually requires heavier analytical and 
computational work \cite{Baldereschi1978,Car1979,Hybertsen1988}.
They have been estimated in various semi-conductors using first-principles
calculations\cite{Resta1981,Resta1983,Baroni1986,Hybertsen1987a,Hybertsen1987} 
and usually renormalize the screening by a few tens of percent. 

The static dielectric function of graphene
has been derived analytically within a 2D Dirac cone model 
\cite{Shung1986,Gorbar2002,Ando2006,Wunsch2006,Barlas2007,
Wang2007a,Hwang2008} in the random phase approximation
(RPA). In those derivations, the role of higher energy electronic
states, the deviation from conical bands  
and the so-called local fields were neglected.
Later, quasiparticle self-consistent GW calculations\cite{Schilfgaarde2011} 
of the screening of point charges in neutral graphene 
seemed to indicate a significant contributions from
the local fields. The general behavior
of the static dielectric function was found to be quite different from 
the analytical RPA derivation. 
However, Coulomb interactions between periodic images were not disabled.
There has been some propositions\cite{Kozinsky2007,Kozinsky2006}, 
within density-functional theory, to correct the contributions from the periodic images.
More simply, complete suppression of those spurious interactions can be achieved
by cutting off the Coulomb interactions between periodic images
\cite{Jarvis1997,Rozzi2006,Ismail-Beigi2006a}.
In a recent study of the energy loss function of neutral isolated graphene 
\cite{Mowbray2014}, the use of a truncated Coulomb interaction (Coulomb cutoff)
was implemented in the framework of time-dependent density-functional theory. 
It was found that the dynamical screening
properties of graphene were strongly affected by the spurious interactions 
between periodic images.

In this work, we focus on the long-wavelength and static screening 
properties of both neutral and doped graphene.
We use density-functional perturbation theory (DFPT) as it includes the 
complete band structure of graphene and the effects of local fields
\cite{Baroni1987,Baroni} and exchange correlation in the local density 
approximation (LDA).
We implement the Coulomb cutoff technique and carefully address the issue of 
extracting two dimensional dielectric properties from 
simulated three-dimensional potentials. 
We then compare our DFPT calculations with the analytical derivations
for the two dimensional massless Dirac fermions within RPA.

In Sec. \ref{sec:Diel_func}, we set the general background of this work by 
defining the static dielectric function in different dimensionality frameworks. 
In Sec. \ref{sec:screening_gr}, we present different methods to calculate 
the static dielectric function of graphene. This includes analytical 
derivations previously developed\cite{Shung1986,Gorbar2002,Ando2006,Wunsch2006,
Barlas2007,Wang2007a,Hwang2008}
and a self-consistent solution implemented in the the phonon package 
of the Quantum ESPRESSO (QE) distribution.
In Sec. \ref{sec:Results} those methods are applied to both doped and neutral 
graphene and the results are compared.

\section{Static dielectric function}
\label{sec:Diel_func}
In this section we introduce the quantities of interest in the formulation 
of the static dielectric response. 
We use the density-functional framework within LDA and atomic units 
to be consistent with the following {\it ab initio} study.
Both the unperturbed system and its response to a perturbative potential are 
described within this framework.
We start with a quick description of the unperturbed system.
Since we are interested only in the static limit here, we consider a 
time-independent Kohn-Sham (KS) potential\cite{Kohn1965}
$V_{\rm{KS}}(\mathbf{r})$, where $\mathbf{r}=(x,y,z)$ is a space variable. 
This potential is the sum of three potentials :
\begin{eqnarray}
V_{\rm{KS}}(\mathbf{r})= V_{\rm{ext}}(\mathbf{r})+ 
V_{\rm{H}}(\mathbf{r})+ V_{\rm{XC}}(\mathbf{r}).
\end{eqnarray}
In the unperturbed system, the external potential 
$V_{\rm{ext}}$ is simply the potential generated by 
the ions of the lattice. 
The remaining potentials are functionals of the electronic density.
The Hartree potential $V_{\rm{H}}$ reads :
\begin{eqnarray}
V_{\rm{H}}(\mathbf{r}) = e^2 \int d\mathbf{r'} 
\frac{n(\mathbf{r'})}{|\mathbf{r}-\mathbf{r'}|},
\end{eqnarray}
and $V_{\rm{XC}}$ is the exchange-correlation potential.  
Since the KS potential determines the solution for the density 
which in turn generates part of the KS potential, 
this approach leads to a self-consistent problem. 
When solved for the system at equilibrium with no perturbation, 
the ground-state density $n_{0}(\mathbf{r})$ is found.

We now proceed to the description of this system within perturbation theory.
An external perturbing potential $\delta V_{\rm{ext}}$ is applied. 
This triggers a perturbation of the electronic density such that the total 
density is
$n_{0}+\delta n$, 
where $\delta n$ is the first order response to the perturbing potential. 
Likewise, all the previously introduced potentials can be 
separated in a equilibrium and perturbed part.
The screened perturbation $\delta V_{\rm{KS}}$ felt by an individual 
electron is the sum of the bare external perturbation $\delta V_{\rm{ext}}$ 
and the screening potential $\delta V_{\rm{H}} + \delta V_{\rm{XC}}$
induced by the density response $\delta n$:
\begin{eqnarray}
\delta V_{\rm{KS}}(\mathbf{r})=
\delta V_{\rm{ext}}(\mathbf{r})+\delta V_{\rm{H}}(\mathbf{r})
+ \delta V_{\rm{XC}}(\mathbf{r}).
\label{eq:potential}
\end{eqnarray}
This leads to an other self-consistent system\cite{Hybertsen1987a} 
solved by the density response $\delta n$. 
From this response we can extract the quantities characterizing 
the screening properties of a material.
The induced electron density $\delta n$ can be seen as independent electrons 
responding to the effective perturbative potential $\delta V_{\rm{KS}}$:
\begin{equation}
\delta n(\mathbf{r})= \int d\mathbf{r'} \chi^0(\mathbf{r},\mathbf{r'}) 
\delta V_{\rm{KS}}(\mathbf{r'}) ,
\label{eq:response}
\end{equation}
thus defining the independent particle static susceptibility  $\chi^0$. 
It can also be seen as interacting electrons responding to the bare external 
perturbative potential:
\begin{equation}
\delta n(\mathbf{r})= \int d\mathbf{r'} \chi(\mathbf{r},\mathbf{r'}) 
\delta V_{\rm{ext}}(\mathbf{r'}) ,
\label{eq:response_int}
\end{equation}
thus defining the interacting particle susceptibility $\chi$. 

We can now proceed to further description of the screening properties of the 
material.
The static dielectric function is first defined in three- and two-dimensional 
frameworks in order to highlight and clarify their differences. 
We then treat the intermediary cases of a 2D-periodic system of finite thickness
and a periodically repeated 2D system, particularly relevant for 
{\it ab-initio} calculations.
For those cases, we will determine the conditions in which it is suitable
to define a 2D static dielectric function.

\subsection{Three-dimensional materials}

In a periodic system, it is more 
convenient to work with the Fourier transform of Eq. \ref{eq:response}. 
Considering a periodic external potential 
$\delta V_{\rm{ext}}(\mathbf{r})=
\delta V_{\rm{ext}}(\mathbf{q}) e^{i\mathbf{q}\cdot \mathbf{r}}$
of wavevector $\mathbf{q}$, we have in linear response theory:
\begin{equation}
\delta n(\mathbf{q}+\mathbf{G})= \sum_{\mathbf{G'}} \chi^0(\mathbf{q}, 
\mathbf{G}, \mathbf{G}') \delta V_{\rm{KS}}(\mathbf{q}+ \mathbf{G}').
\label{eq:chi_0}
\end{equation}
Here, reciprocal lattice wavevectors $\mathbf{G},\mathbf{G'}$ were introduced.
Even though $\delta V_{\rm{ext}}(\mathbf{r})$ only has a $\mathbf{q}$ 
component, the electronic density response can include larger wavevectors 
$\mathbf{q} + \mathbf{G}$.
Consequently, the induced and total potentials can also have 
$\mathbf{q} + \mathbf{G}$ components. Those small wavelength
components (smaller than the lattice periodicity) in the response of the 
electrons are called local fields\cite{Schilfgaarde2011}.  
In a three-dimensional framework, 
the Fourier components of the induced Hartree potential are :
\begin{eqnarray}
\delta V_{\rm{H}}(\mathbf{q} + \mathbf{G}) &=&
\frac{v^{3D}_c(\mathbf{q} + \mathbf{G})}{\kappa_0} 
\delta n(\mathbf{q} + \mathbf{G})
\label{eq:3DPoissonSol}
\\
v^{3D}_c(\mathbf{q} + \mathbf{G})&=&
\frac{4 \pi e^2}{|\mathbf{q} + \mathbf{G}|^2}, 
\label{eq:3DCoul}
\end{eqnarray}
where $v^{3D}_c(\mathbf{q} + \mathbf{G})$ 
is the $\mathbf{q} + \mathbf{G}$ component of the 
Fourier transform of the 3D Coulomb interaction. The static dielectric 
constant $\kappa_0$ renormalizes the Coulomb interaction depending on the 
dielectric environment. We focus here on an isolated graphene layer, 
so that $\kappa_0=1$. 
This constant can also be used in a simple Dirac cone model to include the 
effects of other bands \cite{Shung1986,Gorbar2002,Ando2006,Wunsch2006,
Barlas2007,Wang2007a,Hwang2008}, 
though no definite value has been proposed. 
This will be discussed in Sec. \ref{sec:Results}. Until then, we set 
$\kappa_0=1$.
The Fourier components of the XC potential are written
$\delta V_{\rm{XC}}(\mathbf{q}+\mathbf{G})$.
From Eqs. \ref{eq:potential}, \ref{eq:chi_0} and \ref{eq:3DPoissonSol}, 
we can write :
\begin{eqnarray}
\label{eq:tot_pot}
&&\delta V_{\rm{KS}}(\mathbf{q}+\mathbf{G})=
\delta V_{\rm{ext}}(\mathbf{q}) \  \delta_{\mathbf{G}, \mathbf{0}} + 
\delta V_{\rm{XC}}(\mathbf{q}+\mathbf{G})\\
&& + v^{3D}_c(\mathbf{q} + \mathbf{G}) 
 \sum_{\mathbf{G'}} \chi^0(\mathbf{q} ,\mathbf{G}, \mathbf{G}') 
\delta V_{\rm{KS}}(\mathbf{q}+ \mathbf{G}'), \nonumber
\end{eqnarray}
where $ \delta_{\mathbf{G}, \mathbf{0}}$ represents Kronecker's delta.

The inverse screening function is 
defined as the ratio of the $\mathbf{G}= \mathbf{0}$ component of the 
KS potential (the coarse-grained effective potential) 
over the external potential:
\begin{equation}
\epsilon_{3D}^{-1}(\mathbf{q}) =
\frac{\delta V_{\rm{KS}}(\mathbf{q})}{\delta V_{\rm{ext}}(\mathbf{q})}.
\label{eq:def_eps}
\end{equation}

\subsection{2D materials}
\label{sec:strictly2D}

We now wish to work with 2D electronic densities 
$\delta \tilde{n}(\mathbf{r}_p)$, 
defined in the $\{x,y\}$ plane as follows:
\begin{equation}
\delta\tilde{ n}(\mathbf{r}_p) \equiv
\int_{-\infty}^{+\infty}\delta n(\mathbf{r}_p, z) dz, 
\end{equation}
where $\mathbf{r}_p$ is the in-plane component of $\mathbf{r}$, 
$z$ is the out-of-plane component.

We first consider the system usually studied in analytical derivations,
which will be called the strictly 2D framework.
By strictly 2D, we mean that the electronic density can be written as follows:
\begin{equation}
\delta n(\mathbf{r}_p, z)= \delta\tilde{ n}(\mathbf{r}_p) \delta(z),
\end{equation}
 where $\delta(z)$ is the Dirac delta distribution.
There is no periodicity in the out-of plane direction. 
Considering an external potential  
$\delta V_{\rm{ext}}(\mathbf{q}_p)$
with an in-plane wavevector $\mathbf{q}_p$, we can 
define the Fourier transform of the 2D electronic density 
$\delta \tilde{n}(\mathbf{q}_p+\mathbf{G}_p)$ where $\mathbf{G}_p$ is a 2D 
reciprocal lattice vector. 
The Hartree potential $\delta V_{\rm{H}}(\mathbf{r}_p, z)$ generated by this 
infinitely thin 
electronic distribution is three-dimensional.
We thus separate in-plane and out-of-plane space variables 
to stress the fact that the induced Hartree potential does extend in the 
out-of-plane ($z$) direction, in contrast with the density.
Using Eq. \ref{eq:3DPoissonSol} and performing an inverse Fourier transform in 
the out-of-plane direction only, we find:
\begin{equation}
\delta V_{\rm{H}}(\mathbf{q}_p+\mathbf{G}_p, z)=
\frac{2 \pi e^2}{|\mathbf{q}_p+\mathbf{G}_p|}  
\delta \tilde{n}(\mathbf{q}_p+\mathbf{G}_p)
e^{-|\mathbf{q}_p+\mathbf{G}_p| z} .
\label{eq:Vs_of_delta}
\end{equation}
For our purpose, only the value of the Hartree potential 
where the electrons lie $\delta V_{\rm{H}}(\mathbf{q}_p+\mathbf{G}_p, z=0)$
is of interest. 
Similarly, the KS potential also extends in the out-of-plane direction, but we 
consider only the $z=0$ value. We can work in a 2D reciprocal space with 
$\delta \tilde{n}(\mathbf{q}_p+\mathbf{G}_p)$ and the 
following potentials:
\begin{eqnarray} 
\delta \tilde{V}_{\rm{H}}(\mathbf{q}_p+\mathbf{G}_p) &\equiv&
\delta V_{\rm{H}}(\mathbf{q}_p+\mathbf{G}_p, z=0) \\
\delta \tilde{V}_{\rm{KS}}(\mathbf{q}_p+\mathbf{G}_p) &\equiv&
\delta V_{\rm{KS}}(\mathbf{q}_p+\mathbf{G}_p, z=0) \\
\delta \tilde{V}_{\rm{XC}}(\mathbf{q}_p+\mathbf{G}_p) &\equiv&
\delta V_{\rm{XC}}(\mathbf{q}_p+\mathbf{G}_p, z=0) \\
\delta \tilde{V}_{\rm{ext}}(\mathbf{q}_p) &\equiv&
\delta V_{\rm{ext}}(\mathbf{q}_p, z=0).
\end{eqnarray}
Note that since $\mathbf{q}_p$ is in-plane, 
$ \delta \tilde{V}_{\rm{ext}}(\mathbf{q}_p) =
\delta V_{\rm{ext}}(\mathbf{q}_p, z)$. 
It is then common practice to use the 2D version of  
Eq. \ref{eq:3DPoissonSol}, with the 2D Coulomb interaction 
$v^{2D}_c(\mathbf{q}_p+\mathbf{G}_p)$ (and $\kappa_0=1$) :
\begin{eqnarray}
\delta \tilde{V}_{\rm{H}}(\mathbf{q}_p+\mathbf{G}_p)&=& 
v^{2D}_c(\mathbf{q}_p+\mathbf{G}_p) \delta \tilde{n}(\mathbf{q}_p+\mathbf{G}_p)
\label{eq:2DPoissonSol} 
\\
v^{2D}_c(\mathbf{q}_p+\mathbf{G}_p)&=&
\frac{2 \pi e^2}{|\mathbf{q}_p+\mathbf{G}_p|}.
\end{eqnarray}
We also define a 2D independent particle susceptibility as follows :
\begin{equation}
\delta \tilde{n}(\mathbf{q}_p+\mathbf{G}_p)= 
\sum_{\mathbf{G'}_p} \tilde{\chi}^0(\mathbf{q}, 
\mathbf{G}_p, \mathbf{G}_p') 
\delta \tilde{V}_{\rm{KS}}(\mathbf{q}_p+ \mathbf{G}_p').
\label{eq:chi_02D}
\end{equation}
Working with the 2D quantities defined above, Eq. \ref{eq:tot_pot} becomes :
\begin{eqnarray}
&&\delta \tilde{V}_{\rm{KS}}(\mathbf{q}_p+\mathbf{G}_p)=
\delta \tilde{V}_{\rm{ext}}(\mathbf{q}_p) \  \delta_{\mathbf{G}_p, \mathbf{0}} 
+ 
\delta \tilde{V}_{\rm{XC}}(\mathbf{q}_p+\mathbf{G}_p) \nonumber \\
&&+v^{2D}_c(\mathbf{q}_p + \mathbf{G}_p) 
 \sum_{\mathbf{G}'_p} \tilde{\chi}^0(\mathbf{q}_p ,\mathbf{G}_p, \mathbf{G}'_p) 
\delta \tilde{V}_{\rm{KS}}(\mathbf{q}_p+ \mathbf{G}'_p), 
\label{eq:Vtot_2D}
\end{eqnarray}
and the definition of the inverse screening function is 
modified as follows:
\begin{equation}
\epsilon_{2D}^{-1}(\mathbf{q}_p) =
\frac{\delta \tilde{V}_{\rm{KS}}(\mathbf{q}_p)}
{\delta \tilde{V}_{\rm{ext}}(\mathbf{q}_p)}.
\label{eq:def_eps2D}
\end{equation}
We wish to use this definition in an {\it ab initio} framework. 
This raises some issues that we address now.

\subsection{2D-periodic materials with finite thickness}
\label{sec:Thick_system}

In {\it ab initio} calculations, the electronic density extends also in the 
out-of-plane direction.
In this section we consider the consequences of a finite out-of-plane thickness 
of the electronic density.
We consider now an isolated layer with an electron density of thickness $2d$. 
The results of the purely 2D system should be recovered if the wavelength of 
the perturbation is very large compared to $d$. 
We illustrate this idea by considering an electronic density such that :
\begin{eqnarray}
\label{eq:dn_2Dthick}
\int_{-\infty}^{+\infty} \delta n(\mathbf{r}_p, z) dz &=& 
\delta \tilde{n}(\mathbf{r}_p)\\
\delta n(\mathbf{r}_p, z) = 0 \ \ &\text{if  }& |z|>d \ . \nonumber
\end{eqnarray}
Using Eq. \ref{eq:3DPoissonSol}, the $z=0$ value of the Hartree potential is 
then found to be :
\begin{eqnarray}
\delta \tilde{V}_{\rm{H}}(\mathbf{q}_p+\mathbf{G}_p) &=& 
v_c^{2D}(\mathbf{q}_p+\mathbf{G}_p)   \\  
&\times& 
\int_{-d}^{+d} e^{-|\mathbf{q}_p+\mathbf{G}_p| |z|} 
\delta n(\mathbf{q}_p+\mathbf{G}_p, z) dz. \nonumber
\end{eqnarray}
From this equation one can easily deduce that the condition 
$|\mathbf{q}_p+\mathbf{G}_p| d \ll 1$ is necessary to obtain 
results equivalent to the strictly 2D system. 
Since the largest value of $|\mathbf{G}_p|^{-1}$ is only a fraction of the 
lattice parameter, the above condition can only be fulfilled for 
$\mathbf{G}_p=\mathbf{0}$.
The $\mathbf{q}_p+\mathbf{G}_p$ components of the induced perturbation have 
wavelengths comparable or much smaller than $d$ and the thickness of the 
electronic density cannot be ignored.
However, as long as $|\mathbf{q}_p| d \ll 1$, the coarse-grained induced 
potential can be written:
\begin{equation}
\delta \tilde{V}_{\rm{H}}(\mathbf{q}_p) \approx 
v_c^{2D}(\mathbf{q}_p)  \delta \tilde{n}(\mathbf{q}_p).
\end{equation}
Working with reasonably small perturbation wavevectors, 
the $z=0$ value of the coarse-grained induced potential is equivalent 
to that of the purely 2D system. 
It is then reasonable to use Eq. \ref{eq:Vtot_2D} at $\mathbf{G}_p=\mathbf{0}$
and it makes sense to define the dielectric 
function as in Eq. \ref{eq:def_eps2D}.

\subsection{2D materials periodically repeated in the third dimension}
\label{sec:Perio_images}
In {\it ab initio} calculations, in addition to the non-zero thickness of the 
simulated electronic density, an other issue arises.
Current DFT packages such as QE rely on the use of 3D 
plane waves, requiring the presence of periodic images of the 2D 
system in the out-of-plane direction, separated by 
a distance $c$ (interlayer distance).
For many quantities, imposing a large distance 
between periodic images is sufficient to obtain relevant results for the 2D 
system.
However, simulating the electronic screening of 2D systems correctly is 
computationally challenging due to the long-range character of the Coulomb 
interaction. 
As illustrated in Eq. \ref{eq:Vs_of_delta}, the Hartree potential induced 
by a 2D electronic density perturbed at wavevector $\mathbf{q}_p$ 
goes to zero in the out-of-plane direction on a 
length scale $1/|\mathbf{q}_p|$. For the layers (or periodic images) 
to be effectively isolated, they would have to be separated by a distance 
much greater than $1/|\mathbf{q}_p|$.
The computational cost of calculations increasing linearly with interlayer 
distance, fulfilling this condition for the wavevectors considered in the 
following is extremely challenging.
It is thus preferable to use an alternative method.

In order to isolate the layers from one another, the long-range 
Coulomb interaction is cutoff between layers, as previously proposed 
in such context\cite{Rozzi2006,Ismail-Beigi2006a,Mowbray2014}.
We use the following definition of the Coulomb interaction in real space:
\begin{eqnarray}
\bar{v}_c(\mathbf{r}_p, z) &=& 
\frac{e^2 \theta(l_z-|z|)}{\sqrt{|\mathbf{r}_p|^2+z^2}},
\end{eqnarray}
where $\theta(z)=1$ if $z \ge 0$ and $\theta(z)=0$ if $z < 0$.
The cutoff distance $l_z$ should be small enough that electrons from 
different layers don't see each other, 
but large enough that electrons within the same layer do. 
In other words, if $d$ is representative of the thickness of the electronic 
density, we need the following inequalities to be true :
\begin{eqnarray}
d < l_z < c-d .
\end{eqnarray}
The interlayer distance can be chosen such that $c \gg d$ within reasonable 
computational cost. Then we choose to cutoff the Coulomb potential midway 
between the layers, $l_z=\frac{c}{2}$. 
The Coulomb interaction is generally used in 
reciprocal space. Setting $l_z=\frac{c}{2}$ and considering an external 
perturbative potential with in-plane wavevector 
$\delta V_{\rm{ext}}(\mathbf{q}_p)$,
the Fourier transform of the above 
Coulomb interaction is written as follows\cite{Rozzi2006,Ismail-Beigi2006a}:
\begin{eqnarray}
\bar{v}_c(\mathbf{q}_p+\mathbf{G}_p, G_z) &=&
 \frac{4\pi e^2}{ |\mathbf{q}_p+\mathbf{G}_p|^2+G_z^2} \\
& \times &\left[1 - e^{-|\mathbf{q}_p+\mathbf{G}_p| l_z}  
\cos(G_z l_z)  \right], 
\nonumber 
\end{eqnarray}
where $G_z$ is the out-of-plane component of the reciprocal lattice vector 
$\mathbf{G}$.
In an {\it ab initio} framework, the 3D Coulomb interaction $v_c^{3D}$ should 
thus be replaced by the cutoff Coulomb interaction $\bar{v}_c$:
\begin{equation}
\delta V_{\rm{H}}(\mathbf{q}_p+\mathbf{G}_p, G_z) = 
\bar{v}_c(\mathbf{q}_p+\mathbf{G}_p, G_z)  
\delta n(\mathbf{q}_p+\mathbf{G}_p, G_z).
\end{equation}
Within the DFT LDA framework, the exchange-correlation potential 
is short-range, such that we can neglect interlayer interactions originating 
from that term. 
When the Coulomb interaction is cutoff and within the region $z \in [-l_z;+l_z]$, 
everything happens as if the system was isolated, 
and it can be treated as the 2D-periodic system with finite thickness of the 
previous paragraph.
For the layer at $z=0$, and as long as $|\mathbf{q}_p| d \ll 1$, we can thus 
work with the $z=0$ values of the potentials and use 
the definition of Eq. \ref{eq:def_eps2D} for the dielectric function.

\section{Static screening properties of graphene}
\label{sec:screening_gr}

In this section we present several methods to calculate the inverse static 
dielectric function of graphene. 
First, the derivation of an analytical expression and a semi-numerical 
solution are presented, 
following Refs. \onlinecite{Shung1986,Gorbar2002,Ando2006,Wunsch2006,
Barlas2007,Wang2007a,Hwang2008}.
Graphene is treated as a strictly 2D material, its electronic structure is
represented by the Dirac cone model, the random phase approximation is used 
and local fields are neglected.
Then, we present an {\it ab initio} method based on the phonon package of QE. 
This second method allows one to relax the approximations involved in the 
analytical derivations.

\subsection{Analytical and semi-numerical solutions}
\label{sec:Ana2Dsol}

When the out-of-plane thickness of the electronic density 
can be neglected with respect to the wavelength of the external potential, 
we can work in a strictly 2D framework and
Eqs. \ref{eq:Vtot_2D} and \ref{eq:def_eps2D} can be used. 
In this section, 
two other approximations are used to simplify Eq. \ref{eq:Vtot_2D}. 
Namely, we set 
$\delta \tilde{V}_{\rm{XC}}(\mathbf{q}_p+\mathbf{G}_p)=0$ (RPA) 
and we neglect the local fields, that is, all $\mathbf{G}_p \ne 0$ components.
Eq. \ref{eq:def_eps2D} then reads : 
\begin{equation}
\epsilon_{2D}^{-1}(\mathbf{q}_p) =\frac{1}{1-\frac{2 \pi e^2}{|\mathbf{q}_p|}
\tilde{\chi}^0(\mathbf{q}_p)},
\label{eq:eps_from_chi0}
\end{equation}
where it is understood that 
$\tilde{\chi}^0(\mathbf{q}_p)=
\tilde{\chi}^0(\mathbf{q}_p, \mathbf{0},\mathbf{0})$.
In a model including only $\pi-\pi^*$ bands, 
the independent particle susceptibility is written as follows 
\cite{Shung1986,Gorbar2002,Ando2006,Wunsch2006,
Barlas2007,Wang2007a,Hwang2008}: 
\begin{eqnarray}
\tilde{\chi}^0(\mathbf{q}_p)&=&\frac{1}{\pi^2} 
\int_{\mathbf{K}} d^2 \mathbf{k} 
\sum_{s,s'}|\langle \mathbf{k},s|\mathbf{k}+\mathbf{q}_p,s' \rangle|^2
\frac{f^s_{\mathbf{k}}-f^{s'}_{\mathbf{k}+\mathbf{q}_p}}
{\varepsilon^{s}_{\mathbf{k}}-\varepsilon^{s'}_{\mathbf{k}+\mathbf{q}_p}} .
\nonumber \\
\label{eq:exp_chi0}
\end{eqnarray}
The integral is carried out over electronic wavevectors 
$\mathbf{k}$ in one valley around Dirac point $\mathbf{K}$, 
with a factor two for valley degeneracy. The indexes $s$ and $s'$ designate 
the $\pi$ or $\pi^*$ bands.
The occupation of the state of momentum $\mathbf{k}$ in band $s$ is labelled 
$f^{s}_{\mathbf{k}}$ and $\varepsilon^{s}_{\mathbf{k}}$ is the corresponding 
energy.
Within the Dirac cone model, a linear dispersion is assumed 
$\varepsilon^{s}_{\mathbf{k}}=s \ \hbar v_F |\mathbf{k}|$, 
with $s=-1$ ($s=+1$) for the $\pi$ ($\pi^*$) band, and $v_F$ is the Fermi 
velocity. 
The wavefunctions overlap is then written 
$ |\langle \mathbf{k},s|\mathbf{k+q},s' \rangle|^2= 
(1+ss'\cos(\theta_{\mathbf{k}}-\theta_{\mathbf{k}+\mathbf{q}_p}))/2$, 
where $\theta_{\mathbf{k}}$ ($\theta_{\mathbf{k}+\mathbf{q}_p}$)
is the angle between $\mathbf{k}$ ($\mathbf{k}+\mathbf{q}_p$) and 
an arbitrary reference axis. The Dirac cone band structure is isotropic and 
$\tilde{\chi}^0$ depends only on the norm of the perturbation wavevector 
$|\mathbf{q}_p|$.
The numerical implementation of this integral in the Dirac cone model
will be referred to as "semi-numerical solution". 
It has the advantage of accounting for temperature effects.
In the zero temperature limit and following the tedious but straightforward 
calculus in Refs. \onlinecite{Shung1986,Gorbar2002,Ando2006,Wunsch2006,
Barlas2007,Wang2007a,Hwang2008}, 
the following analytical forms can 
be found. In the case $|\mathbf{q}_p| \le 2k_F$:  
\begin{eqnarray}
\epsilon_{2D}(|\mathbf{q}_p|) &=& 1+\frac{2e^2}{ \hbar v_F} 
\frac{2 k_F}{|\mathbf{q}_p|}, 
\label{eq:Ana_0K1}
\end{eqnarray}
where $k_F=\frac{|\varepsilon_F|}{\hbar v_F}$ is the Fermi wavevector, if 
$\varepsilon_F$ is the Fermi energy taken from the Dirac point.
In the case $|\mathbf{q}_p| > 2k_F$:
\begin{eqnarray}
\label{eq:Ana_0K2}
\epsilon_{2D}(|\mathbf{q}_p|) &=& 1+\frac{2 e^2}{ \hbar v_F} 
\frac{2 k_F}{|\mathbf{q}_p|} \times \Bigg[  \frac{\pi |\mathbf{q}_p| }{8 k_F} + 
\\ 
 && 1-\frac{1}{2} 
\sqrt{1-\frac{4k_F^2}{|\mathbf{q}_p|^2}}-\frac{|\mathbf{q}_p|}{4 k_F}
\sin^{-1}\left(\frac{2k_F}{|\mathbf{q}_p|} \right) \Bigg] . \nonumber
\end{eqnarray}
Those expressions are relevant for doped graphene. For neutral graphene, we are 
in the case $|\mathbf{q}_p| > 2k_F$, 
but since $k_F \to 0$, Eq. \ref{eq:Ana_0K2} simplifies to :
\begin{eqnarray}
\label{eq:Ana_0Kneutral}
\epsilon_{2D}(|\mathbf{q}_p|) &=& 1+\frac{ \pi e^2}{ 2\hbar v_F}.  
\end{eqnarray}
The following work aims at investigating the validity of those expressions.

\subsection{ DFPT LDA solution}
\label{sec:DFPT_eps}

Several approximations (Dirac cone model, neglecting local fields, RPA...)
were used in order to derive the previous analytical expressions.
Their validity is not obvious in graphene. 
{\it Ab initio} methods like DFPT offer the opportunity to 
relax those approximations \cite{Hybertsen1987a}.
In this section we detail how we obtain the 2D static dielectric function 
as defined in Eq. \ref{eq:def_eps2D} from DFPT.  
The issues of the periodic images and finite thickness in the out-of plane 
direction are treated as previously discussed.
The remaining issues are to apply the adequate perturbation 
and extract relevant 2D quantities.
The equilibrium system is calculated using the usual DFT plane-waves package. 
At that point, interlayer interactions can be neglected in graphene. 
To study the screening properties, we develop
the response of the electronic density to an external potential within QE. 
The code originally calculates the induced electronic density in response to a 
phonon perturbation\cite{Baroni}. 
Here, we replace the phonon perturbation 
by the perturbation $\delta V_{\rm{ext}}(\mathbf{q}_p)$. 
This perturbation is constant in the 
out-of-plane direction and modulated by a single wavevector $\mathbf{q}_p$
in the plane.
As shown previously, the relevant quantity is the $z=0$ value of the 
KS potential, coarse-grained in the plane 
$\delta \tilde{V}_{\rm{KS}}(\mathbf{q}_p)$. 
Note that the $G_z \ne 0$ components are needed to perform a 
Fourier transform and then take the $z=0$ value.
%We print $\delta V_{\rm{KS}}(\mathbf{q}_p, G_z)$ and calculate :
%\begin{eqnarray}
%\delta \tilde{V}_{\rm{KS}}(\mathbf{q}_p) = 
%\sum_{G_z} \delta V_{\rm{KS}}(\mathbf{q}_p, G_z).
%\end{eqnarray}
The number of $G_z$ elements is limited only 
by the kinetic energy cutoff.
We then use the definition of Eq. \ref{eq:def_eps2D}.

\subsubsection{Technical details of DFPT calculations}
\label{sec:tech_details}
Our DFT/DFPT calculations were performed using the Quantum ESPRESSO 
distribution\cite{Giannozzi2009}. The electronic structure is obtained by DFT 
calculations within the local density approximation\cite{Perdew1981} (LDA).
Since the electronic structure is calculated without cutoff, 
it can contain some spurious interlayer states above the Dirac point. 
In the calculations, it is thus safer to dope graphene with holes to avoid 
those states. 
We will assume electron-hole symmetry and consider the 
following results valid for both electron and hole doping.
We use norm-conserving pseudo-potentials with 2s and 2p states in valence and 
cutoff radii of $0.78$ \AA.
We use a $0.01$ Ry Methfessel-Paxton smearing function for the electronic 
integrations, a $65$ Ry kinetic energy cutoff, and a $96\times 96 \times 1$ 
electron-momentum grid. 
The lattice parameter is $a=2.46$ \AA\ and the distance between 
graphene and its periodic images is $c = 4.0 \times a \approx 9.8$ \AA.
The Coulomb interaction is cutoff when calculating
the response of the system to an external perturbative potential.
The results presented here were obtained for a perturbation wavevector in the 
direction $\mathbf{\Gamma} \to \mathbf{K}$ of the Brillouin zone. 
Identical calculations were performed in different directions. 
The variations on the results were small enough
to assume that the screening properties of graphene are isotropic. 
Occasionally, variations from this setup were required and will be specified.

\subsubsection{Valididity of the 2D framework}
Now we quickly discuss the validity the 2D treatment with respect to the 
thickness of the electronic density.
\begin{figure}[h]
\includegraphics[width=0.48\textwidth]{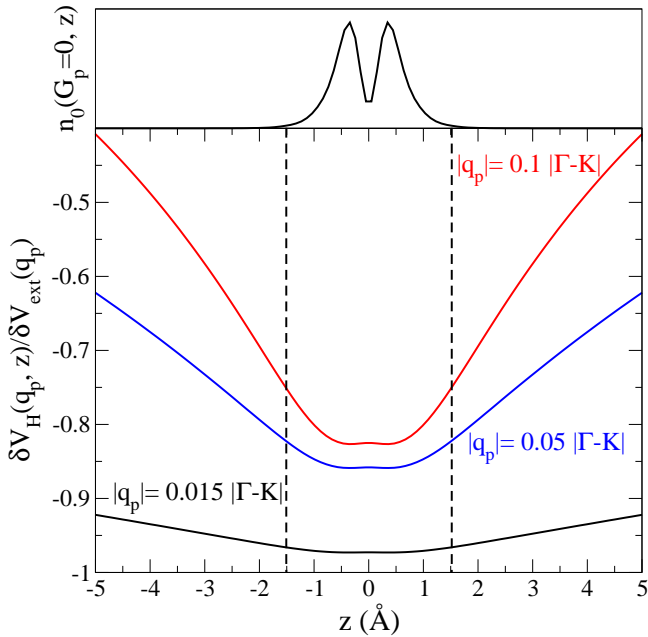}
\caption{(Color online) The induced potential 
$\delta V_{\rm{H}}(\mathbf{q}_p, z)$ in the out-of-plane direction 
at different values of $|\mathbf{q}_p|$, expressed in units of the distance 
between the $\mathbf{\Gamma}$ and $\mathbf{K}$ points of the Brillouin zone. 
Calculations were performed for 
a Fermi level of $\varepsilon_F=0.25$ eV, taken from the Dirac point. 
Details of the numerical calculations can be found in 
Sec. \ref{sec:tech_details}. 
The typical profile of $n_0(\mathbf{G}_p=\mathbf{0},z)$ is represented. 
The equilibrium density was chosen here to have a common reference for all 
perturbations.}
\label{fig:thickness}
\end{figure}
Fig. \ref{fig:thickness} shows the out-of plane variations of the 
coarse-grained induced potential $\delta V_{\rm{H}}(\mathbf{q}_p,z)$ and the 
equilibrium electronic density $n_0(\mathbf{G}_p=\mathbf{0},z)$ of  a single 
isolated graphene layer in our {\it ab initio} framework. 
We use three values of $|\mathbf{q}_p|$ 
covering the range of values used in the following section. 
In that range, Fig. \ref{fig:thickness} shows negligible variations of the
the induced potential over the extent of the electron distribution. 
The two-dimensional description of the screening properties is thus valid.
This range of wavevectors covers a large span of situations where 
static screening plays a role.
For example, in the case of electronic transport we are typically interested in 
values of $|\mathbf{q}_p|$ on the scale of the Fermi wavevector 
for relatively small doping levels. Thickness effects are negligible in this
situation.

\section{Results}
\label{sec:Results}

In this section we present the results of the full DFPT LDA method 
(Sec. \ref{sec:DFPT_eps}, labelled "LDA") 
and compare them to the analytical solution 
(Eqs. \ref{eq:Ana_0K1}-\ref{eq:Ana_0Kneutral}, labelled "Analytical") 
for the static dielectric function of doped and neutral graphene.
We identify the contributions of temperature, bands, local fields, and 
exchange-correlation by using different methods.
When the analytical derivation presented in Sec. \ref{sec:Ana2Dsol} is used, 
the Fermi velocity is the only parameter needed to define the Dirac cone band 
structure. 
For consistency with the {\it ab initio} methods, we use the Fermi velocity 
obtained in the linear part of the DFT band structure, such that 
$\hbar v_F= 5.49$ eV$\cdot$\AA. 
It is well known that electron-electron interactions increase this value by 
approximately $20\%$ (depending on doping) within the GW 
approximation\cite{Attaccalite2010}. 
The renormalized value is in good agreement with experiments.
This renormalization is ignored here, but should be accounted for 
when comparing with experiment.
Three intermediary methods were used to investigate the differences 
between the analytical solution and the self-consistent DFPT LDA solution. 
The first is the semi-numerical method introduced in Sec. \ref{sec:Ana2Dsol}. 
The independent particle susceptibility $\tilde{\chi}^0(\mathbf{q}_p)$ is 
obtained by numerical integration of Eq. \ref{eq:exp_chi0}, and inserted into 
Eq. \ref{eq:eps_from_chi0}. This solution relies on the same approximations 
as the analytical solution but it can be carried out at a chosen temperature 
(or energy smearing) as long 
as the integration grid is adequately fine.
The second is labelled "RPA" and consists in setting the exchange correlation 
potential to zero within the DFPT method. 
The third is labelled "RPA no LF" and consists in evaluating
the DFPT independent particle susceptibility and 
inserting it in Eq. \ref{eq:eps_from_chi0}.
This implies using RPA and neglecting local fields, 
as well as a strictly 2D treatment,
since Eq. \ref{eq:eps_from_chi0} was derived in a strictly 2D framework. 
This method boils down to the evaluation of Eq. \ref{eq:eps_from_chi0}, 
within a more complete {\it ab initio} model for the band structure.
Table \ref{tab:methods} summarizes the labels and main characteristics of the 
various methods used in the following plots.
\begin{table}[h]
\caption{Summary of the various methods used in the plots of 
Sec. \ref{sec:Results}.
For each method, we report : 
i) the treatment of electron-electron correlation, LDA refering to the 
use of the XC potential within LDA, "$=0$" meaning that the XC potential is set 
to zero ; 
ii) wether local fields are included (YES) or neglected (NO) ; and 
iii) which band structure model was used, the full {\it ab initio} band 
structure or the simpler Dirac cone model for $\pi-\pi^*$ bands.}
\begin{tabular}{ l c c c }
\hline
Label       &  Exchange-Correlation     & Local Fields     & Bands         \\
\hline 
LDA         &      LDA        & YES             & {\it ab initio}  \\
RPA         &      $=0$       & YES             & {\it ab initio}   \\
RPA no LF   &      $=0$       & NO             &  {\it ab initio}  \\
Analytical  &      $=0$       &  NO             & Dirac cones   \\ 
\hline
\end{tabular}
\label{tab:methods}
\end{table}

\subsection{Importance of cutting off the Coulomb interactions}
\begin{figure}[h!]
\centering
\includegraphics[width=0.48\textwidth]{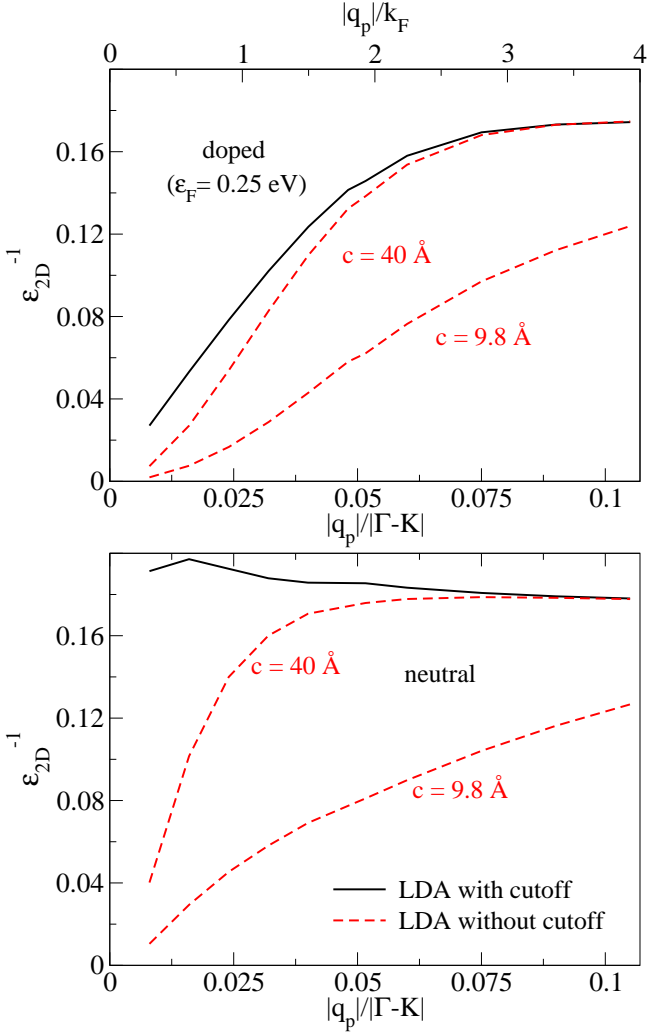}
\caption{(Color online) DFPT LDA results are plotted with and 
without cutoff of the Coulomb interactions.
The inverse dielectric function, as defined in Eq. \ref{eq:def_eps2D}, 
is plotted
as a function of the adimensional variable 
$|\mathbf{q}_p|/|\mathbf{\Gamma} -\mathbf{K}|$, 
where $|\mathbf{\Gamma} -\mathbf{K}| \approx 1.7$ \AA$^{-1}$ is the distance 
between the $\mathbf{\Gamma}$ and $\mathbf{K}$ points of the Brillouin zone. 
The calculations were performed for neutral graphene (lower panel) 
and doped graphene (upper panel) with $\varepsilon_F=0.25$ eV, 
measured with respect to the Dirac point. In the upper panel, we also represent 
the scale $|\mathbf{q}_p|/k_F$ where 
$k_F \approx 0.27 |\mathbf{\Gamma} -\mathbf{K}|$ refers to the Fermi wavevector 
in the doped case.
Two interlayer distances were used $c \approx 40$ \AA\ and $c \approx 9.8$ \AA\ 
to convey the dependency of the results (without cutoff) on that parameter.
When the Coulomb interaction is cutoff, the results are independent of 
the interlayer distance $c$. Finally, note that for the neutral case at 
small wavevectors and with cutoff, the results are quite sensitive 
to energy smearing/grid effects. In this situation, we used a 
$140 \times 140 \times 1$ grid and $0.005$ Ry energy smearing to be as close as 
room temperature as manageable.}
\label{fig:eps2D-3D}
\end{figure}
\begin{figure}[h]
\centering
\includegraphics[width=0.48\textwidth]{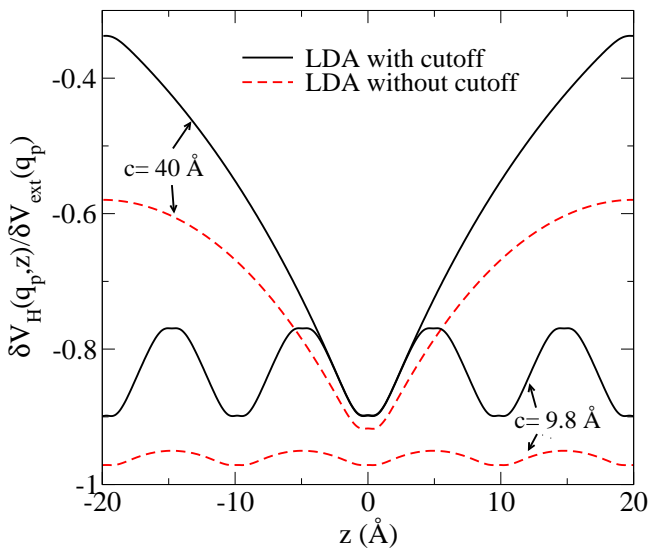}
\caption{(Color online) The Hartree potentials are plotted in the out-of-plane 
direction with and without cutoff and for two different interlayer distances 
$c \approx 40$ \AA\ and $c \approx 9.8$ \AA\ .
The calculations were performed 
for doped graphene with $\varepsilon_F=0.25$ eV, at 
$|\mathbf{q}_p|\approx 0.32 \ |\mathbf{\Gamma} -\mathbf{K}| \approx 1.2\ k_F$.}
\label{fig:pot_cutoff}
\end{figure}

We begin by presenting the DFPT LDA results and pointing 
out the importance of
the Coulomb cutoff in Fig. \ref{fig:eps2D-3D}.
We plot the inverse dielectric function obtained with the LDA method with and 
without cutoff. 
In the latter case, we follow the process of  Sec. \ref{sec:DFPT_eps} but the 
original 3D Coulomb interaction $v^{3D}_c$ is used. 
Two different interlayer distances are displayed, 
namely $c \approx 9.8$ \AA\ and $c \approx 40$ \AA\ . 
It is clear that interlayer interactions play a major role in the screening 
without cutoff, as a strong dependency on the interlayer distance is shown.
For $c\approx 9.8$, the effect of the cutoff is drastic.
When the interlayer distance is increased, 
the results without cutoff slowly approach the results with cutoff. 
This is also the case in the limit of large wavevector. 
In general, the results with and without cutoff are similar when the scale 
on which the induced Hartree potential decreases $1/|\mathbf{q}_p|$
is negligible compared to the interlayer distance $c$.
However, even using large interlayer distance, the effect of
cutting off the Coulomb interactions remains significant.
To obtain accurate {\it ab initio} results for an isolated layer, 
it is thus essential to cutoff the Coulomb interactions.
To give a clearer picture of the effects of the Coulomb cutoff, 
we plot the Hartree potential with and without cutoff for two different 
interlayer distances $c$ in Fig. \ref{fig:pot_cutoff}.
With cutoff, the Hartree potentials corresponding to the two interlayer 
distances coincide exactly with each other within the region $[-l_z;+l_z]$, 
$l_z$ being half the smaller interlayer distance here.
This confirms that within this region, everything happens as if the layers were 
isolated.
Without cutoff, in contrast, the Hartree potentials are significantly 
different, stressing the fundamental difference in the response of systems with 
different interlayer distances.

\subsection{Comparison of analytical and LDA methods: band structure effects}

\begin{figure}[h]
\centering
\includegraphics[width=0.48\textwidth]{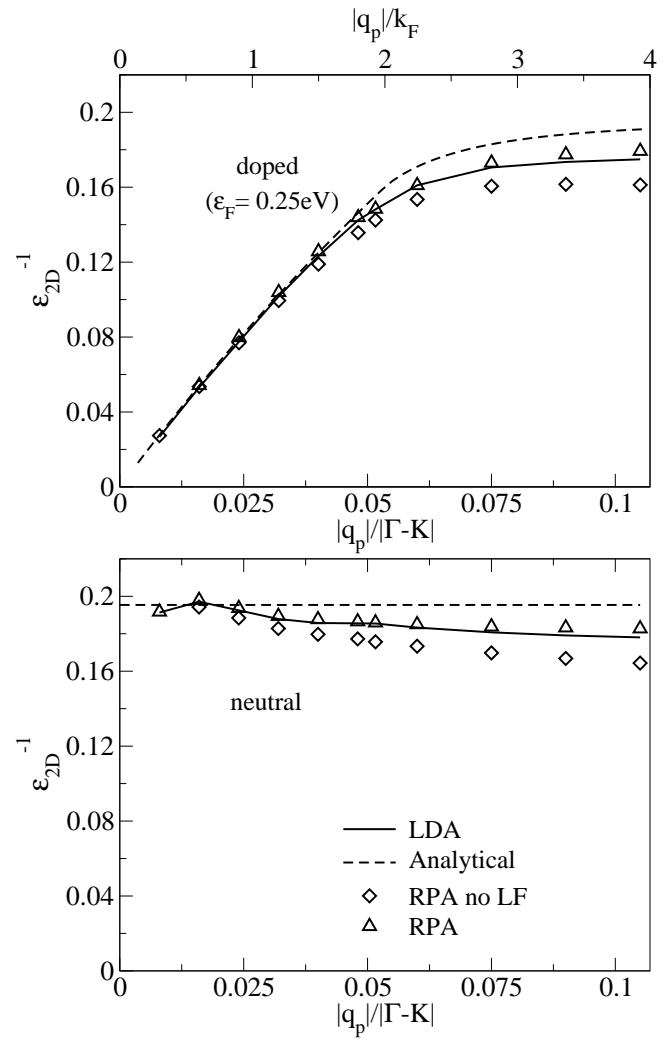}
\caption{ Comparison of the static dielectric function of graphene 
obtained within the  LDA and analytical methods. 
We use the same axes as in Fig. \ref{fig:eps2D-3D}.
We also plot the results of the "RPA" and "RPA no LF" methods.
For this last method applied to neutral graphene, 
the point with the smallest $|\mathbf{q}_p|$ was not converged and is not 
represented.}
\label{fig:eps2D-Ana}
\end{figure}

In Fig. \ref{fig:eps2D-Ana}, we compare the LDA results (with cutoff)
to the analytical solution of Eqs. \ref{eq:Ana_0K1}-\ref{eq:Ana_0Kneutral}.
The results of the two methods are rather close overall. 
In doped graphene, the LDA results are in very good agreement ($\approx 3\%$)
with the analytical method for $|\mathbf{q}_p| \le 2 k_F$.
A more pronounced discrepancy ($\approx 10 \%$) is observed for 
$|\mathbf{q}_p| >  2 k_F$.
In the neutral case, a similar $\approx 10 \%$ discrepancy occurs for most 
values of $|\mathbf{q}_p|$, but agreement seems to be reached in the small    
$|\mathbf{q}_p|$ limit. For neutral graphene at small wavevectors, 
smearing plays a significant role.
Though not plotted here, the semi-numerical method is 
equivalent to the analytical solution when performed with an energy 
smearing corresponding to room temperature. 
Using the same energy smearing and grid  
as in DFPT to perform the numerical integration of Eq. \ref{eq:exp_chi0}
showed that smearing effects are negligible except in the small wavevector limit
of the neutral case. 
For DFPT LDA calculations in this regime, we lowered the smearing to $0.005$ Ry 
and changed the grid accordingly to $140\times 140 \times 1$ in Figs. 
\ref{fig:eps2D-3D} and \ref{fig:eps2D-Ana}. For this smearing, 
agreement between LDA and analytical results is reached around 
$|\mathbf{q}_p| \approx 0.025 |\mathbf{\Gamma}-\mathbf{K}|$.
Although quite low in terms of what is computationally manageable in DFT, 
this energy smearing is still large compared to the value corresponding to 
room temperature.
At room temperature, we expect that DFPT LDA calculations would show
the agreement to be reached for smaller $|\mathbf{q}_p|$. 
In the $0$ temperature limit, it should be reached for $|\mathbf{q}_p| \to 0$.
Thus, for graphene in general, we can consider that
LDA and analytical results significantly differ only for 
$|\mathbf{q}_p| > 2 k_F$, which corresponds to $|\mathbf{q}_p| > 0$ 
in the neutral case.

To investigate the origin the $\approx 10\%$ discrepancy above $2k_F$, 
we use the aforementioned "RPA no LF" method. 
In Fig. \ref{fig:eps2D-Ana}, this method gives 
a smaller inverse dielectric constant than both the LDA ($\approx 8\%$) 
and analytical ($\approx 16\%$) methods above $2k_F$.
Comparing the "RPA no LF" and LDA methods indicates that
the combined effect of RPA, neglecting local fields, and a strictly 2D framework
is a $\approx 8\%$ decrease of the results.
As mentioned before, the band structure model is the only difference between
the "RPA no LF" and analytical methods. This suggests that
the effects of using the Dirac cone approximation are more sizable 
($\approx 16\%$) but somewhat compensate the other approximations. 
Overall, we end up with the $\approx 10\%$ discrepancy above $2k_F$ 
between LDA and analytical method.
When setting the exchange correlation potential to zero in DFPT, 
see "RPA" in Fig.\ref{fig:eps2D-Ana}, the results are only slightly changed.
This means that neglecting the local fields in the plane 
(what is meant by RPA in the derivation of Eq. \ref{eq:eps_from_chi0}) 
and out-of-plane
(equivalent to making the strictly 2D approximation) have more important 
effects than exchange-correlation. 
Although the use of an LDA exchange-correlation potential has negligible 
consequences for the results presented here, 
we would like to point out that such potentials are derived in the framework of 
a three-dimensional electron gas. 
Consequently, their relevance in a 2D framework is limited and the RPA
method might be more reliable than the LDA one.

\begin{figure}[h]
\centering
\includegraphics[width=0.48\textwidth]{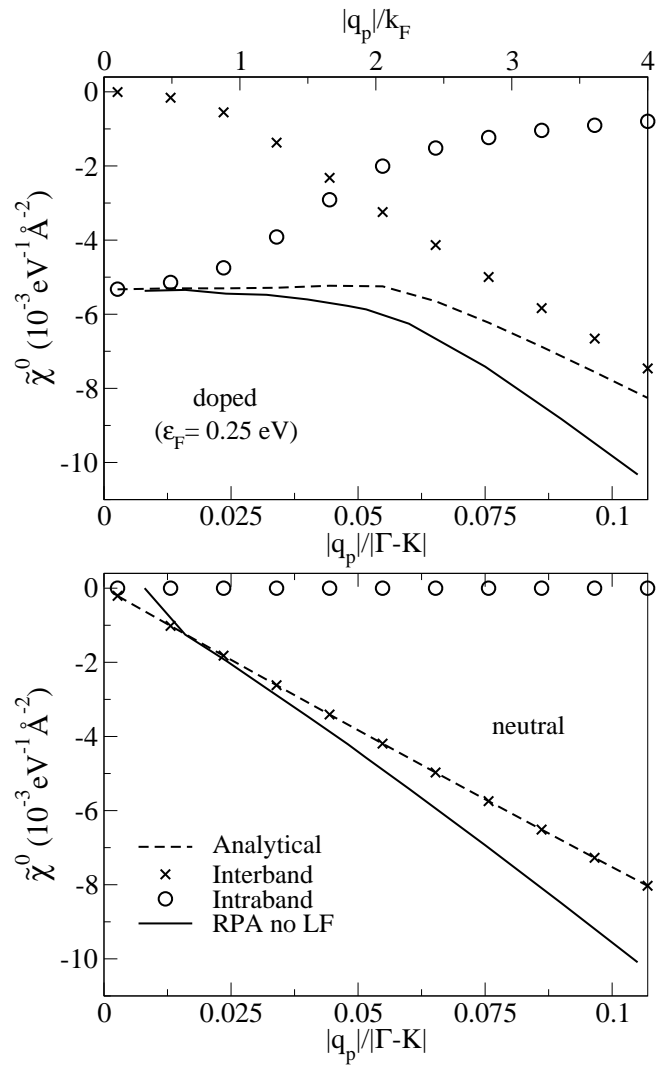}
\caption{Comparison of the independent electron susceptibility of graphene 
obtained within the "RPA no LF" and analytical methods. 
We use the same axes as in Fig. \ref{fig:eps2D-3D}.
The contribution of intraband and interband processes to $\tilde{\chi}^0$
are represented by circles and crosses, respectively.
To calculate those contributions, 
we used the semi-numerical solution with a small energy smearing ($0.001$ Ry). 
The analytical and semi-numerical methods 
are equivalent in that case.}
\label{fig:chi0}
\end{figure}

A better interpretation of the effects of band structure can be achieved by 
comparison of the independent particle susceptibility $\tilde{\chi}^0$ from the 
"RPA no LF" and analytical methods in Fig. \ref{fig:chi0}.
In the $|\mathbf{q}_p| \le 2 k_F$ regime, 
the screening is dominated by the zeroth-order of 
$\tilde{\chi}^0$, proportional to the density of states. 
The linear part of the DFT band structure of graphene is well represented by 
the Dirac cone model. 
As long as the Fermi level is reasonably small (but finite), 
the densities of states obtained in DFT and analytically are very close.
We then find a very good agreement with the analytical 
derivation in this regime. 
In the upper panel of Fig. \ref{fig:chi0}, it is clear that a higher-order 
(in $|\mathbf{q}_p|$ ) term
in $\tilde{\chi}^0$ from DFPT is responsible for the gradual disagreement with 
the analytical solution as $|\mathbf{q}_p|$ increases.
In the neutral case, the zeroth order of $\tilde{\chi}^0$ vanishes
with the density of state, and $\tilde{\chi}^0$ is always dominated by 
contributions of higher-order terms.
For the $|\mathbf{q}_p| > 2 k_F$ regime in general, 
the first-order in $|\mathbf{q}_p|$ seems to dominate. 
The susceptibility $\tilde{\chi}^0$ is then ruled by interband processes, 
some of them going beyond the range of validity of the Dirac cone model. 

Overall, we find a rather good agreement with the analytical derivation of 
Refs. \onlinecite{Shung1986,Gorbar2002,Ando2006,Wunsch2006,
Barlas2007,Wang2007a,Hwang2008}. 
This is in strong contrast with the conclusions of a previous {\it ab-initio} 
study\cite{Schilfgaarde2011} of the screening of point charges in neutral 
graphene.
Our work differs notably on the use of a Coulomb cutoff, and the treatment of 
{\it ab initio} results to extract the 2D screening properties of a system that 
is effectively 3D. 
The authors of Ref. \onlinecite{Schilfgaarde2011} state that they 
checked the negligibility of the interlayer interactions by looking at the effects 
of interlayer distance on the bands. Such test is misleading. 
Indeed, interlayer interactions are negligible on the bands for spacing 
larger than $\approx 5$ \AA. However,  as discussed in Sec. \ref{sec:Perio_images}, 
the interlayer interactions affect the calculation of the dielectric response 
when the wavelength of the perturbation is comparable with the interlayer distance, 
making the use of a Coulomb cutoff essential.
We can also comment on the use of the constant $\kappa_0$ 
in Eq. \ref{eq:3DPoissonSol} 
to include the effects of other bands. 
Such a constant is not appropriate since it 
would affect all the orders in $\tilde{\chi}^0$, 
including the zeroth order that is correct. 
To have an analytic expression quantitatively closer to the DFPT LDA results, 
one should only renormalize the contribution from the interband processes.
Finally, as mentioned before, we used the DFT Fermi velocity in this work.
One should keep in mind that within the GW approximation and consistent with
experimental results, the Fermi velocity is increased by $20 \%$. 
This yields very similar curves, with a $\approx 16 \%$ increase of the 
value of $\epsilon^{-1}_{2D}$ at large $\mathbf{q}_p$, as easily found 
by plotting the analytical expressions.

\section{Conclusion}
Definitions of the dielectric function depend on the dimensionality. 
The study of the screening properties of 2D materials first requires 
precise definitions of the relevant quantities. 
After setting such a formalism, we review previous analytical derivations
of the screening properties of graphene. We highlight the approximations
involved in those derivations and propose a DFT-based method to overcome them.
The DFPT method with Coulomb cutoff presented here 
is general and can be applied to study the screening properties of other 2D 
materials.
We showed that cutting off the Coulomb interactions 
is essential to recover the screening properties of an isolated layer.
Our DFPT LDA calculations on graphene lead to an inverse dielectric 
function that is very close to the analytical form of 
Refs. \onlinecite{Shung1986,Gorbar2002,Ando2006,Wunsch2006,
Barlas2007,Wang2007a,Hwang2008} for 
$|\mathbf{q}_p| \le 2 k_F$, and smaller by $\approx 10\%$ for 
$|\mathbf{q}_p| > 2 k_F$. 
Overall, the Dirac cone model in a strictly 2D framework, 
in the zero temperature limit, using RPA 
and neglecting local fields leads to a quite accurate and 
simple analytical expression for the static 
dielectric function of graphene. 
Smearing effects are negligible at room temperature and 
exchange-correlation effects within LDA are also quite small.
Neglecting the local-fields leads to a $\approx 8\%$ 
underestimation of the inverse dielectric function above $2k_F$.
The largest error comes from the Dirac cone model for the band structure.
This model remains an excellent approximation in the 
$|\mathbf{q}_p| \le 2k_F$ regime, as long as the Fermi level lies 
in the region where the bands are linear. In the $|\mathbf{q}_p|>2k_F$ regime, 
however, the Dirac cone model leads to a $\approx 16\%$ overestimation of the 
inverse dielectric function due to the contribution of interband processes 
probing states beyond the Dirac cones.
This overestimation compensates the local fields effects and the analytical 
model ends up overestimating the DFPT LDA inverse dielectric function by 
$\approx 10\%$ above $2k_F$.

\begin{acknowledgments}
The authors acknowledge support from the Graphene Flagship and
from the French state funds managed by
the ANR within the Investissements d'Avenir programme under reference
ANR-13-IS10- 0003-01.
This work was granted access to the HPC resources of The Institute for scientific
Computing and Simulation financed by Region Ile de France and the project Equip@Meso
(reference ANR-10-EQPX- 29-01) overseen by the French National Research Agency (ANR)
as part of the "Investissements d'Avenir" program.
Computer facilities were also provided by CINES, CCRT and IDRIS
(project no. x2014091202).
\end{acknowledgments}

%% ****** End of file apssamp.tex ******
\bibliographystyle{apsrev4-1}
\bibliography{Screening}

\end{document}